# Optimal client recommendation for market makers in illiquid financial products

Dieter Hendricks* and Stephen J. Roberts

Machine Learning Research Group, Oxford-Man Institute of Quantitative Finance
Department of Engineering Science, University of Oxford

**Abstract.** The process of liquidity provision in financial markets can result in prolonged exposure to illiquid instruments for market makers. In this case, where a proprietary position is not desired, pro-actively targeting the *right* client who is likely to be interested can be an effective means to offset this position, rather than relying on commensurate interest arising through natural demand. In this paper, we consider the inference of a client profile for the purpose of corporate bond recommendation, based on typical recorded information available to the market maker. Given a historical record of corporate bond transactions and bond meta-data, we use a topic-modelling analogy to develop a probabilistic technique for compiling a curated list of client recommendations for a particular bond that needs to be traded, ranked by probability of interest. We show that a model based on Latent Dirichlet Allocation offers promising performance to deliver relevant recommendations for sales traders.

## 1 Introduction

The exchange of financial products primarily relies on the principle of matching willing counter-parties who have opposing interest in the underlying product, resulting in a demand-driven natural transaction at an agreed price. There are, however, cases where there is insufficient commensurate demand on one side at the desired price level, resulting in one of the parties needing to either wait for willing counter-parties or adjust their price. Where transaction immediacy is required, the client may approach a *market maker* (such as a bank or broker) who will facilitate the required trade by guaranteeing the other side of the transaction and charging a fee (the *spread*) for this service. This process of facilitating client transactions is termed *liquidity provision*, as the client can pay a fee to trade an otherwise illiquid asset immediately.

From the market maker's perspective, providing this liquidity of course results in taking a proprietary position in the underlying product, affecting their inventory and/or cash on hand. The management of this inventory and how it relates to quoted spread to account for associated risks is widely studied (see [2, 8, 11, 12] as examples), however is beyond the scope of this paper. We are interested in the particular case where a market maker has provided liquidity in a product and is not interested in a long-term proprietary position, viz. they would like to mitigate or eliminate this exposure by targeting interested clients to offset the position. Finding suitable clients is the task of *sales traders*, who use their knowledge of potential clients' interests to find a match for the required trade, however understanding the nuanced preferences of all the clients is an arduous task. This paper seeks to create a system which will automate client profile inference and assist the sales traders by providing them with a curated list of clients to contact, who are most likely to be interested in the product. A successful system will expedite the liquidation of the market maker's product exposure, assisting with regulatory [9, 16] and inventory management [1] concerns.

---
\* Corresponding author: dieter.hendricks@eng.ox.ac.uk

The products we consider are *corporate bonds*, which are fixed-term financial instruments issued by companies as a means of raising capital for operations. An investor who owns a corporate bond is usually entitled to interest payments from the issuer in the form of regular *coupons*, and redemption of the *face value* of the bond at *maturity*. The *yield* (interest rate) associated with a corporate bond is typically higher than a comparable government-issued bond. This yield differential is commensurate with the perceived credit-worthiness of the underlying company, the nature of the issue (senior/subordinated, secured/unsecured, callable/non-callable, etc.), the liquidity of the market place and the contractual provisions for contingencies in the event of issuer default [17, 10]. From an investors perspective, corporate bonds offer a relatively stable investment compared to, say, buying stocks in the company, since the instrument does not participate in the underlying profits of the company and bondholders are preferential creditors in the case of company bankruptcy. Following the initial issuance, corporate bonds are traded between investors in the *secondary market* until maturity, where market makers facilitate transactions by providing liquidity when required, leading to product exposures which need to be offset, as discussed above.

We will use a topic modelling analogy to frame the problem and develop a client profile inference technique. In the Natural Language Processing (NLP) community, many authors have focused on probabilistic generative models for text corpora, which infer latent statistical structure in groups of documents to reveal likely topic attributions for words [19, 6, 5, 24]. One such model is Latent Dirichlet Allocation (LDA) [6], which is a three-level hierarchical Bayesian model under which *documents* are modelled as a finite mixture of *topics*, and topics in turn are modelled as a finite mixture over *words* in the vocabulary. Learning the relevant topic mixture and word mixture probabilities provides an explicit statistical representation for each document in the corpus. If one considers *documents* as *products* and *words* as *clients*, this has a natural analogy to the client recommendation problem we seek to solve. By observing product-client (document-word) transactions, we can infer a posterior probability of trade over *relevant* clients (topic with highest probability mass) for a particular product. These ideas are made more concrete in Section 2. Sampling from this posterior probability distribution provides us with a mechanism for client recommenda-tion (most likely matches), coupled with a probability of trade, which will assist sales trades to gauge recommendation confidence.

This paper proceeds as follows: Section 2 discusses the analogy between topic modelling and bond recommendation. Section 3 introduces LDA as a candidate technique for client profile inference. Section 4 discusses some baseline models for comparison. Section 5 introduces some custom metrics to quantify recommendation efficacy, in the context of bond recommendation. Section 6 discusses the data and results, and Section 7 provides some concluding remarks.

## 2 A topic modelling approach: Terminology and analogies

We will frame the problem using the exposition in Blei et al. [6] as a guide, making appropriate modifications to reflect the bond recommendation use-case.

The *word* ($w$) represents the basic observable unit of discrete data, where each word belongs to a finite vocabulary set indexed by $\{1, ..., W\}$. Where appropriate, we may use the convention of a superscript ($w^i$) to indicate location in a sequence (such as in a document or topic), and subscript ($w_t$) to indicate a word observed at a particular time. Words are typically represented using unit-basis $W$-length vectors, with a 1 coinciding with the associated vocabulary index and zeros elsewhere. In our context, words represent *clients*, viz. $w = i$ is a unit vector associated with client $i$. We have used the term *client interest*, as we may abstract the actual trade status of our recorded data (traded, not traded, indication of interest, traded away, passed) to an indicator representing *interest* or *no interest*. In each case, the client was interested in the underlying bond and requested a price, regardless of whether they actually traded with us, another bank or changed their mind. This is the behaviour we would like to predict and has the added benefit of reducing the sparsity of our dataset. In future work, we may consider relaxing this assumption to determine if certain trade statuses contain more relevant information for likely client interest.

A *document* ($d$) is a sequence of $N$ words $d = \{w^1, w^2, ..., w^N\}$, where $w^n$ is the $n^{th}$ word in the sequence. In our context, a document relates to a specific *product*, where, like a *document* is a collection of *words*, a *product* represents a collection of *clients* who have expressed interest to trade.

A *topic* ($z$) is a collection of $M$ words $z = \{w^1, w^2, ..., w^M\}$ which are related in some way, representing an abstraction of words which can act as a basic building block of documents. In our context, a topic refers to a *client group*, viz. a set of clients that are regarded as *similar* based on the products they are interested in.

A *corpus* ($\mathbf{w}$) is a collection of $D$ documents, $\mathbf{w} = \{d^1, d^2, ..., d^D\}$. In our context, the corpus represents the set of *products* which the market maker is interested in trading with its clients.

### 2.1 The product-client term-frequency matrix

In the topic modelling analogy, a corpus can be summarised by a *document-word matrix*, which is essentially a 2-d matrix where, for each document (row), we count the frequency of each possible word in the vocabulary (columns) in the document. This summary is justified by the *exchangeability* assumption typical in topic modelling, where temporal and spatial ordering of documents and words are ignored to ensure tractable inference.

For our application, we can compute an analogous *product-client matrix* where, for each product (row), we count the number of times a client (column) has expressed interest in the product. While we suspect the temporal property of client interest is an important property (clients trade bonds in response to particular market conditions, to renew exposure close to maturity or as part of a regular portfolio rebalancing scheme), we will ignore these effects in this study and revisit these properties in future work. We will, however, ensure only *active* bonds are used to populate the product-client matrix, i.e. bonds which have a start date before the training period start and maturity date after the chosen testing day.

The product-client summary of records we use in this study results in a highly sparse matrix, with relatively few clients dominating trading activity. Since equal weight is placed on zero and non-zero counts, this will make inference for clients who trade less frequently more difficult. One remedy used in the topic modelling literature is to convert the raw document-word matrix into a Term Frequency-Inverse Document Frequency (TF-IDF) matrix [23, 21]. Under this scheme, for our application, the weighting of a client associated with a product increases proportionally with the number of times they have traded the product, but this is offset by the number of times the product is traded among all clients. We will use the standard formulation,

$$\text{tf-idf}(w, d, \mathbf{w}) = \text{tf}(w, d) \cdot \text{idf}(w, \mathbf{w}), \tag{1}$$

where

$$\text{tf}(w, d) = 0.5 + 0.5 \cdot \frac{f_{w,d}}{\max\{f_{w^*,d} : w^* \in d\}}$$

and

$$\text{idf}(w, \mathbf{w}) = \log \frac{D}{|\{d \in \mathbf{w} : w \in d\}|}.$$

Here, $f_{w,d}$ is the raw count of the number of times client $w$ was interested in product $d$, $D$ is the total number of products and $\mathbf{w}$ is the set of all products.

## 3 Latent Dirichlet Allocation

Latent Dirichlet Allocation (LDA) [6] is a probabilistic generative model typically used in Natural Language Processing (NLP) to infer latent topics present in sets of documents. Documents are modelled as a mixture of topics sampled from a Dirichlet prior distribution, where each topic, in turn, corresponds to a multinomial distribution over words in the vocabulary [13]. The learned document-topic and topic-word distributions can then be used to identify the best topics which describe the document, as well as the best words which describe the associated topics [7].

As discussed in Section 2, we will consider *documents* as *products* and *words* as *clients*, allowing us to infer a posterior probability of trade (or at least client interest) over *relevant* clients (topic with highest probability mass) for a particular product.

LDA is traditionally a *bag-of-words* model, assuming document and word *exchangeability*. This means an entire corpus is used to infer document-topic and topic-word distributions, ignoring potential effects of spatial and temporal ordering. Given the particular problem of corporate bond recommendation, certain spatial and temporal features may be useful for more accurate recommendations. For example, the maturity date and frequency of coupon payment associated with a particular bond may influence the client's probability of trading. The duration and convexity characteristics of a bond and it's impact on the client's overall exposures may affect their willingness to trade. In this paper, we will ignore the effects of bond characteristics and temporal ordering of transactions, using only the bond *issue* and *maturity* dates to ensure they are *active* for the training and testing periods.

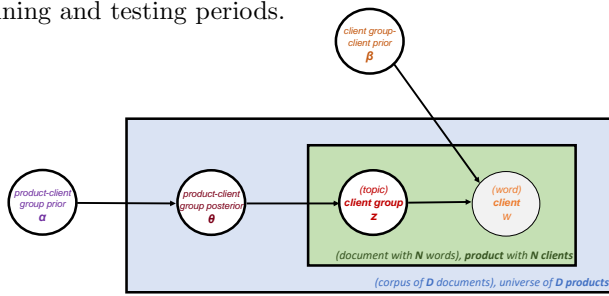

**Fig. 1.** Graphical representation of LDA in plate notation, indicating interpretation of *words*, *topics* and *documents* as *clients*, *client groups* and *products*.

To formalise ideas, we will reproduce the key aspects of the mathematical exposition of LDA (we follow conventions and notation set out in Wallach [24]), modified to reflect the product recommendation use-case. This is complemented by the *plate notation* representation of LDA in Figure 1.

Client generation is defined by the conditional distribution $P(w_t = i | z_t = k)$, described by $T(W - 1)$ free parameters, where $T$ is the *number of client groups* and $W$ is the *total number of clients*. These parameters are denoted by $\Phi$, with $P(w_t = i | z_t = k) \equiv \phi_{i|k}$. The $k^{th}$ row of $\Phi$ ($\phi_k$) thus contains the distribution over *clients* for *client group* $k$.

Client group generation is defined by the conditional distribution $P(z_t = k | d_t = d)$, described by $D(T-1)$ free parameters, where $D$ is the *total number of products traded by the market maker*. These parameters are denoted by $\Theta$, with $P(z_t = k | d_t = d) \equiv \theta_{k|d}$. The $d^{th}$ row of $\Theta$ ($\theta_d$) thus contains the distribution over *client groups* for *product* $d$.

The joint probability of a *set of products* $\mathbf{w}$ and a set of associated latent *groups of interested clients* $\mathbf{z}$ is

$$P(\mathbf{w}, \mathbf{z} | \Phi, \Theta) = \prod_i \prod_k \prod_d \phi_{i|k}^{N_{i|k}} \theta_{k|d}^{N_{k|d}}, \qquad (2)$$

where $N_{i|k}$ is the number of times *client i* has been generated by *client group k*, and $N_{k|d}$ is the number of times *client group k* has been interested in *product d*.

As in Blei et al. [6], we assume a Dirichlet prior over $\boldsymbol{\Phi}$ and $\boldsymbol{\Theta}$, i.e.

$$P(\boldsymbol{\Phi}|\beta\mathbf{m}) = \prod_k \mathrm{Dirichlet}(\phi_k|\beta\mathbf{m}) \quad (3)$$

and

$$P(\boldsymbol{\Theta}|\alpha\mathbf{n}) = \prod_d \mathrm{Dirichlet}(\theta_d|\alpha\mathbf{n}). \quad (4)$$

Combining these priors with Equation 2 and integrating over $\boldsymbol{\Phi}$ and $\boldsymbol{\Theta}$ yields the probability of the *set of products* given hyperparameters $\alpha\mathbf{n}$ and $\beta\mathbf{m}$:

$$P(\mathbf{w}|\alpha\mathbf{n},\beta\mathbf{m}) = \sum_{\mathbf{z}} \Big( \prod_k \frac{\prod_i \Gamma(N_{i|k} + \beta m_i)}{\Gamma(N_k) + \beta} \frac{\Gamma(\beta)}{\prod_i \Gamma(\beta m_i)} \\ \prod_d \frac{\prod_k \Gamma(N_{k|d} + \alpha n_k)}{\Gamma(N_d + \alpha)} \frac{\Gamma(\alpha)}{\prod_k \Gamma(\alpha n_k)} \Big). \quad (5)$$

In Equation 5, $N_k$ is the total number of times *client group k* occurs in $\mathbf{z}$ and $N_d$ is the number of *clients* interested in *product d*. This posterior is intractable for exact inference, but a number of approximation schemes have been developed, notably Markov Chain Monte Carlo (MCMC) [15] and variational approximation [13, 14].

For our study, we made use of the *scikit-learn* [20] open-source Python library, which includes an implementation of the online variational Bayes algorithm for LDA, described in Hoffman et al. [13, 14]. They make use of a simpler, tractable distribution to approximate Equation 5, optimising the associated variational parameters to maximise the Evidence Lower Bound (ELBO), and hence minimising the Kullback-Leibler (KL) divergence between the approximating distribution and the true posterior.

## 4 Baseline models for comparison

1. *Empirical Term-Frequency (ETF)*: We can use the normalised product-client term-frequency matrix discussed in Section 2.1 to construct an empirical probability distribution over clients for each product. This encodes the historical intensities of client interest, without exploiting any latent structure.

2. *Non-negative Matrix Factorisation (NMF)*: NMF aims to discover latent structure in a given non-negative matrix by using the product of two low-rank non-negative matrices as an approximation to the original, and minimising the distance of the reconstruction to the original, measured by the Frobenius norm [18]. Applied to our problem, for a specified number of client groups, NMF can be used to reveal an *unnormalised* probability distribution over client groups for each product, and distribution over clients for each client group, from a given term-frequency matrix. These probabilities can be normalised for comparison with other models.

## 5 Evaluating recommendation efficacy

Recommender systems are usually evaluated in terms of their predictive accuracy, but the appropriate metrics should be chosen to reflect success in the specific application [22]. The data we have for inference and testing purposes is framed in terms of *positive interest*, viz. the presence of a record indicates a client was interest in the associated product, and the absence of a record indicates no interest. In addition, we are interested in capturing the accuracy of a "top $N$" client list, as opposed to a binary classifier. In terms of the standard confusion matrix metrics, we will thus focus on true and false positive results, however we have implemented a nuanced interpretation based on our application:

- *Cumulative True Positives (CTP)*: A client recommendation for a particular product is classified as a True Positive (TP) if the recommended client matches the actual client for that product on the testing day. The total number of TPs for a testing day is thus the total number of correctly matched recommendations. Given our use-case, where the $N$ best (ranked) recommendations are sampled, we compute the *cumulative* TPs as the number of TPs captured within the first $x$ recommendations, $x = 1, ..., N$. More formally, the CTP for product $j$ captured within the first $x$ recommendations is given by

$$\text{CTP}_j^x = \sum_{i=1}^{x} \mathbb{I}_{(w_j^i = w_j^*)}, \tag{6}$$

  where $w_j^i$ is the $i^{th}$ recommended client for product $j$ and $w_j^*$ is the actual client who traded product $j$.

- *Relevant False Positives (RFP)*: A client recommendation is classified as a Relevant False Positive (RFP) if is does not match the actual client for that product on that day, but the recommended client traded another product instead. The rationale here is that the model captures the property of general client trading interest, so may be useful for the sales traders to discuss possibilities with the client, even though the model has matched the client to the incorrect product. These are measured at the first recommendation level ($x = 1$). For product $j$,

$$\text{RFP}_j = \mathbb{I}_{\left((w_j^1 \neq w_j^*) \cap (w_j^1 \in \{w_k^*\}_{k \neq j})\right)}. \tag{7}$$

- *Irrelevant False Positives (IFP)*: A client recommendation is classified as an Irrelevant False Positive (IFP) if is does not match the actual client for that product on that day, and the recommended client did not trade another product. This captures the *wasted resources* property of a false positive, as the sales trader could have spent that time targeting the right client. These are measured at the first recommendation level ($x = 1$). For product $j$,

$$\text{IFP}_j = \mathbb{I}_{\left((w_j^1 \neq w_j^*) \cap (w_j^1 \notin \{w_k^*\}_{k \neq j})\right)}. \tag{8}$$

## 6 Data and results

*Data:* BNP Paribas (BNPP) provided daily recorded transactions with clients for various corporate bond products over the period 5 January 2015 to 10 February 2017, including records where clients did not end up trading with the bank. To maintain privacy, the Client and Product ID's were anonymised in the provided dataset. The data includes the following fields:

- *TradeDateKey*: Date of the transaction (*yyyymmdd*)
- *VoiceElec*: Whether the transaction was performed over the phone (*VOICE*) or electronically (*ELEC, ELECDONE*)
- *BuySell*: The trade direction of the transaction
- *NotionalEUR*: The notional of the bond transaction, in EUR
- *Seniority*: The seniority of the bond
- *Currency*: The currency of the actual transaction
- *TradeStatus*: Indicates whether the bond was actually traded with the bank (*Done*), price requested but traded with another LP (*TradedAway*), bank decided to pass on the trade (*Passed*), client requested price without immediate intention to trade (*IOI*) or client did not end up trading (*TimeOut, NotTraded*). This field also refers to entries which are aggregate bond positions based on quarterly reports (*IPREO*). Some entries also indicate an *UNKNOWN* trade status.
- *IsinIdx*: The unique product ID associated with the bond.
- *ClientIdx*: The unique client ID.

Some metadata was also provided, related to properties of the traded bonds:

- *Currency*: Currency of the bond
- *Seniority*: Seniority of the bond
- *ActualMaturityDateKey*: Maturity date of bond (*yyyymmdd*)
- *IssueDateKey*: Issue date of bond (*yyyymmdd*)
- *Mat*: Maturity as number of days since "00 Jan 2000" (*00000100*)
- *IssueMat*: Issue date as number of days since "00 Jan 2000" (*00000100*)
- *IsinIdx*: Unique product ID associated with bond
- *TickerIdx*: Bond type index

This data was parsed by: 1) Removing *TradeStatus = IPREO* or *UNKNOWN*, 2) Collapsing the *TradeStatus* column into a single *client interest* indicator, 3) Isolating either *Buys* or *Sells* for inference related to a particular trade direction, 4) Ensuring bonds are "active" for the relevant period, i.e. issued before start of training and matures after testing date, and finally, 5) construct a product-client term frequency matrix as described in Section 2.1.

*Results:* Due to space constraints, we will only show results for the SELL trade direction, however results for BUYS were quite similar. Figure 2 shows $CTP^x$ for $x = 1, ..., 100$ for a number of candidate models, with parameter inference from a single training period (5 Jan 2015 to 30 Nov 2016) and model testing on a single period (1 Dec 2016 to 10 Feb 2017). A crude baseline which all models beat is random client sampling (without replacement), indicated by the solid black line, suggesting that there is useful information in the historical transaction record for the purpose of client recommendation. The ETF model does surprisingly well, capturing 40% TP matches within the first 20 recommendations. We find that the LDA models offer superior accuracy beyond 10 recommendations, indicating that the latent structure is useful for the purpose of refining posterior probability of trade. These results do, however, aggregate results over the entire testing period, whereas the intended use-case will be on a daily basis, using the previous day's transactions to refine recommendations.

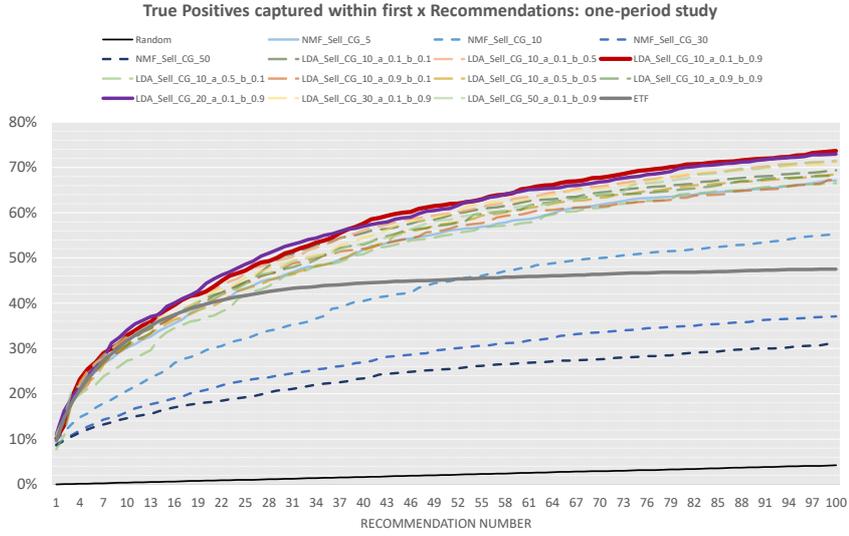

**Fig. 2.** Comparison of candidate models for *single period* training (5 Jan 2015 to 30 Nov 2016) and testing (1 Dec 2016 to 10 Feb 2017), evaluating cumulative true positives captured within first $x$ recommendations. Client SELL interest.

**Table 1.** Summarised results for *through-time* study, varying estimation windows and hyperparameter values. Averaged over testing days in period 05 Jan 2015 to 10 Feb 2017. *WS*=Inference Window Size and *CG*=Client Groups. Client SELL interest.

| Model | WS | CG | $\alpha$ | $\beta$ | $CTP^1$ | $CTP^2$ | $CTP^3$ | $CTP^4$ | $CTP^5$ | $CTP^6$ | $CTP^7$ | $CTP^8$ | $CTP^9$ | $CTP^{10}$ | $\sigma(CTP^{10})$ | RFP | IFP |
|---|---|---|---|---|---|---|---|---|---|---|---|---|---|---|---|---|---|
| ETF | 100 | | | | 0.11 | 0.17 | 0.21 | 0.24 | 0.27 | 0.29 | 0.31 | 0.33 | 0.34 | 0.36 | 0.06 | 0.49 | 0.40 |
| NMF | 100 | 5 | | | 0.10 | 0.14 | 0.16 | 0.19 | 0.21 | 0.23 | 0.24 | 0.26 | 0.27 | 0.28 | 0.07 | 0.83 | 0.07 |
| NMF | 100 | 10 | | | 0.10 | 0.13 | 0.14 | 0.16 | 0.17 | 0.19 | 0.20 | 0.21 | 0.22 | 0.23 | 0.06 | 0.77 | 0.13 |
| NMF | 100 | 20 | | | 0.11 | 0.13 | 0.14 | 0.15 | 0.16 | 0.17 | 0.18 | 0.19 | 0.20 | 0.21 | 0.05 | 0.72 | 0.17 |
| NMF | 100 | 50 | | | 0.10 | 0.12 | 0.13 | 0.14 | 0.14 | 0.15 | 0.15 | 0.16 | 0.16 | 0.17 | 0.04 | 0.63 | 0.27 |
| LDA | 100 | 5 | 0.1 | 0.9 | 0.10 | 0.14 | 0.17 | 0.20 | 0.23 | 0.25 | 0.27 | 0.29 | 0.31 | 0.32 | 0.08 | 0.84 | 0.06 |
| LDA | 100 | 10 | 0.1 | 0.9 | 0.10 | 0.14 | 0.18 | 0.20 | 0.23 | 0.25 | 0.27 | 0.29 | 0.30 | 0.32 | 0.08 | 0.81 | 0.09 |
| LDA | 100 | 20 | 0.1 | 0.9 | 0.10 | 0.15 | 0.18 | 0.21 | 0.23 | 0.25 | 0.27 | 0.29 | 0.31 | 0.32 | 0.08 | 0.79 | 0.11 |
| LDA | 100 | 50 | 0.1 | 0.9 | 0.11 | 0.15 | 0.18 | 0.21 | 0.23 | 0.25 | 0.27 | 0.28 | 0.30 | 0.31 | 0.07 | 0.77 | 0.12 |
| ETF | 500 | | | | 0.11 | 0.17 | 0.22 | 0.25 | 0.27 | 0.30 | 0.32 | 0.34 | 0.36 | 0.38 | 0.07 | 0.54 | 0.35 |
| NMF | 500 | 5 | | | 0.11 | 0.16 | 0.18 | 0.21 | 0.23 | 0.25 | 0.27 | 0.28 | 0.30 | 0.31 | 0.09 | 0.80 | 0.09 |
| NMF | 500 | 10 | | | 0.10 | 0.13 | 0.14 | 0.16 | 0.16 | 0.18 | 0.19 | 0.20 | 0.21 | 0.22 | 0.06 | 0.78 | 0.12 |
| NMF | 500 | 20 | | | 0.11 | 0.12 | 0.14 | 0.15 | 0.16 | 0.17 | 0.18 | 0.19 | 0.20 | 0.21 | 0.05 | 0.73 | 0.17 |
| NMF | 500 | 50 | | | 0.10 | 0.11 | 0.12 | 0.13 | 0.14 | 0.15 | 0.15 | 0.16 | 0.17 | 0.17 | 0.06 | 0.65 | 0.23 |
| LDA | 500 | 5 | 0.1 | 0.9 | 0.11 | 0.16 | 0.21 | 0.22 | 0.24 | 0.26 | 0.28 | 0.30 | 0.33 | 0.34 | 0.10 | 0.82 | 0.06 |
| LDA | 500 | 10 | 0.1 | 0.9 | 0.11 | 0.16 | 0.20 | 0.23 | 0.24 | 0.27 | 0.30 | 0.31 | 0.33 | 0.34 | 0.09 | 0.81 | 0.09 |
| LDA | 500 | 20 | 0.1 | 0.9 | 0.11 | 0.16 | 0.21 | 0.22 | 0.25 | 0.27 | 0.29 | 0.30 | 0.32 | 0.35 | 0.09 | 0.79 | 0.10 |
| LDA | 500 | 50 | 0.1 | 0.9 | 0.12 | 0.17 | 0.20 | 0.23 | 0.25 | 0.27 | 0.29 | 0.30 | 0.32 | 0.34 | 0.09 | 0.79 | 0.11 |

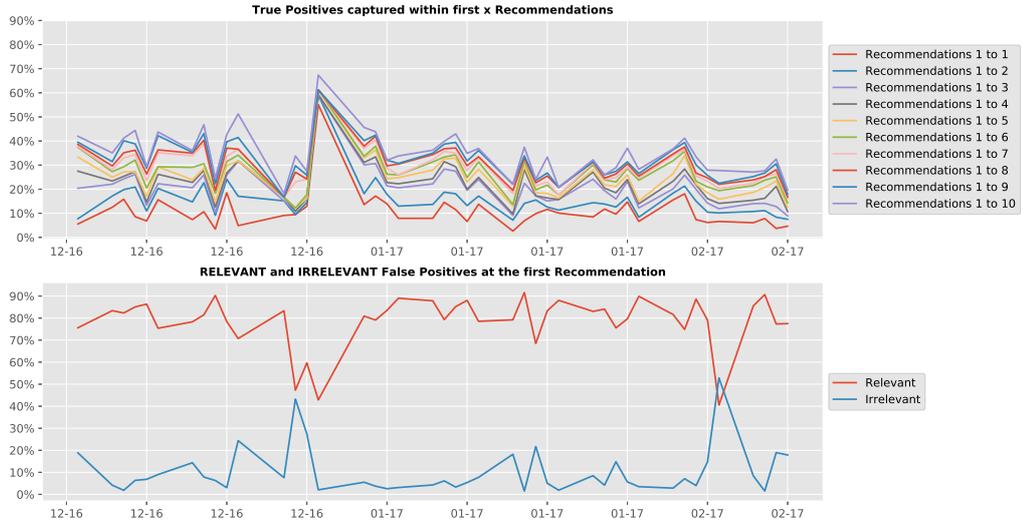

**Fig. 3.** True Positives, Relevant and Irrelevant False Positives. LDA with $CG=20$, $\alpha = 0.1$, $\beta = 0.9$, 500-day rolling training window, 05 Jan 2015 to 10 Feb 2017. Client SELL interest.

Table 1 shows the results from a *through-time* study, where a specified *window size* (WS) (number of days) was used for parameter inference, test metrics calculated for the day after, and the study moved forward one day. Results shown are averaged over all the testing days in the data set. Here, it is clear that, while the ETF model offers comparable CTP results to other models, it offers poor RFP and IFP results. For the highlighted LDA model, on average, 79% of the "incorrectly" recommended clients still traded on that day, albeit a different product. For a sales trader, making contact with these clients could start the conversation about their interests and be converted into a trade. Although it may not solve the direct problem of offsetting a particular position, it could still translate into revenue for the market maker. We found that increasing the WS to 500 days alleviates the sparse data problem somewhat and offers marginal improvements in performance, however more sophisticated data balancing techniques [3] should be explored to ensure accurate inference for clients who trade less frequently.

Figure 3 shows $CTP^x$ for $x = 1, ..., 10$, RFP and IFP for each testing day, using the highlighted *through-time* LDA model in Table 1 (WS = 500, CG = 20, $\alpha = 0.1$, $\beta = 0.9$). We see that this model offers relatively consistent recom-mendation performance. There is a significant increase in CTP accuracy around the end of December 2016, but this is largely due to relatively few "typical" clients trading. These clients would have traded frequently in the past, thus are more likely to be recommended in the first instance. There is also a decrease in performance around the beginning of February 2017. This could be due to a change in client preferences due the expiry of a certain class of bonds. This does suggest that simple moving inference windows may insufficient to capture temporal trends, and a more sophisticated modelling approach may be required.

## 7 Conclusion

We proposed a novel perspective for framing financial product recommendation using a topic modelling analogy. By considering *documents* as *products* and *words* as *clients*, we can use classical NLP techniques to develop a probabilistic generative model to infer an explicit statistical representation for each *product* as a mixture of *client groups* (topics), where each *client group* is a mixture of *clients*. By observing product-client (document-word) transactions, we can infer a posterior probability of trade over *relevant* clients (topic with highest probability mass) for a particular product.

We find that LDA is a promising technique to infer statistical structure from a historical record of client transactions, for the purpose of client recommendation. While it does not necessarily outperform a naïve approach in terms of "top 10" true positive recommendations, it does offer superior "top 100" accuracy and *relevant* false positive performance, where recommended clients trade other products which could translate into revenue for the market maker.

Further research should consider the advantages of inference using balanced product-client term frequency matrices [3], incorporating bond metadata information into the LDA algorithm [25], considering the effects of trends and other temporal phenomena [7], and more sophisticated hierarchical topic modelling techniques to exploit latent structure [5, 4].

*Acknowledgements:* The authors thank BNP Paribas Global Markets for the financial support and provision of data necessary for this study. The discussions with Joe Bonnaud, Laurent Carlier, Julien Dinh, Steven Butlin and Philippe Amzelek provided meaningful context and intuition for the problem.

## References

1. Y. Amihud and H. Mendelson. Asset pricing and the bid-ask spread. *Journal of Financial Economics*, 17(2):223–249, 1986.
2. M. Avellaneda and S. Stoikov. High-frequency trading in a limit order book. *Quantitative Finance*, 8(3):217–224, 2016.
3. G.E. Batista, R.C. Prati, and M.C. Monard. A study of the behavior of several methods for balancing machine learning training data. *ACM Sigkdd Explorations Newsletter*, 6(1):20–29, 2004.

4. D.M. Blei, T.L. Griffiths, and M.I. Jordan. The nested Chinese Restaurant Process and Bayesian nonparametric inference of topic hierarchies. *Journal of the ACM (JACM)*, 57(2):7, 2010.
5. D.M. Blei, T.L. Griffiths, M.I. Jordan, and J.B. Tenenbaum. Hierarchical topic models and the Nested Chinese Restaurant Process. *Advances in Neural Information Processing*, 2004.
6. D.M. Blei, A.Y. Ng, and M.I. Jordan. Latent Dirichlet Allocation. *Journal of Machine Learning Research*, 3(Jan):993–1022, 2003.
7. L. Bolelli, S. Ertekin, and C.L. Giles. Topic and trend detection in text collections using latent dirichlet allocation. *Proceedings of the 31th European Conference on IR Research on Advances in Information Retrieval*, pages 776–780, 2009.
8. S. Das and M. Magdon-Ismail. Adapting to a market shock: Optimal sequential market-making. *Advances in Neural Information Processing Systems*, 2009.
9. Darrell Duffie. Market making under the proposed volcker rule. *Rock Center for Corporate Governance at Stanford University Working Paper No. 106*, 2012.
10. E.J. Elton, M.J. Gruber, D. Agrawal, and C. Mann. Explaining the rate spread on corporate bonds. *The Journal of Finance*, 56(1):247–277, 2001.
11. S. Ghoshal and S. Roberts. Optimal FX market making under inventory risk and adverse selection constraints. *Working paper*, 2016.
12. O. Guéant. Optimal market making. *arXiv:1605.01862 [q-fin.TR]*, 2017.
13. M. Hoffman, F.R. Bach, and D.M. Blei. Online learning for Latent Dirichlet Allocation. *Advances in Neural Information Processing*, 2010.
14. M. Hoffman, D.M. Blei, C. Wang, and J.W. Paisley. Stochastic variational inference. *Journal of Machine Learning Research*, 14(1):1303–1347, 2013.
15. M.I. Jordan. *Learning in graphical models*, volume 89. Springer Science & Business Media, 1998.
16. W.A. Kaal. *Global Encyclopedia of Public Administration, Public Policy, and Governance: Dodd-Frank Act*. Springer International Publishing, 2016.
17. I. Kim, K. Ramaswamy, and S. Sundaresan. The valuation of corporate fixed income securities. *Manuscript, Columbia University*, 1988.
18. D.D. Lee and H.S. Seung. Learning the parts of objects by non-negative matrix factorization. *Nature*, 401:788–791, 1999.
19. D.J.C. MacKay and L.C. Bauman Peto. A hierarchical Dirichlet language model. *Natural Language Engineering*, 1(3):289–308, 1995.
20. F. Pedregosa, G. Varoquaux, A. Gramfort, V. Michel, B. Thirion, O. Grisel, M. Blondel, P. Prettenhofer, R. Weiss, V. Dubourg, et al. Scikit-learn: Machine learning in python. *Journal of Machine Learning Research*, 12(Oct):2825–2830, 2011.
21. S. Robertson. Understanding inverse document frequency: on theoretical arguments for idf. *Journal of Documentation*, 60(5):503–520, 2004.
22. G. Shani and A. Gunawardana. *Evaluating recommendation systems, Recommender systems handbook*. Springer US, 2011.
23. K. Sparck Jones. A statistical interpretation of term specificity and its application in retrieval. *Journal of Documentation*, 28:11–21, 1972.
24. H.M. Wallach. Topic modeling: Beyond bag-of-words. *Proceedings of the 23rd international conference on Machine learning*, pages 977–984, 2006.
25. X. Wang and E. Grimson. Spatial latent dirichlet allocation. *Advances in Neural Information Processing Systems*, pages 1577–1584, 2008.